\RequirePackage[colorlinks = true, urlcolor   = blue, citecolor  = blue]{hyperref}
\documentclass[draft]{agujournal2019}
\usepackage{url} %this package should fix any errors with URLs in refs.
\usepackage{lineno}
\draftfalse

\usepackage{amsmath}
\usepackage{calrsfs}
\DeclareMathOperator{\ra}{\mathit{Ra}}

\usepackage{color}

\usepackage{graphicx}
\usepackage{tikz}
\usetikzlibrary{positioning} % Optional, but useful for node positioning
\usepackage{xr}

\journalname{Geophysical Research Letters}

\begin{document}

\title{Simulation and modelling of convective mixing of carbon dioxide in geological formations}
\authors{Marco De Paoli\affil{1,2}, Francesco Zonta\affil{3}, Lea Enzenberger\affil{2}, Eliza Coliban\affil{2}, and Sergio Pirozzoli\affil{4}}

\affiliation{1}{Physics of Fluids Group and Max Planck Center for Complex Fluid Dynamics and J. M. Burgers Centre for Fluid Dynamics, University of Twente, P.O. Box 217, 7500AE Enschede, The Netherlands}
\affiliation{2}{Institute of Fluid Mechanics and Heat Transfer, TU Wien, 1060 Vienna, Austria}
\affiliation{3}{School of Engineering, Newcastle University, Newcastle upon Tyne NE1 7RU, United Kingdom}
\affiliation{4}{Dipartimento di Ingegneria Meccanica e Aerospaziale, Sapienza Universit\`a di Roma, 00184 Rome, Italy}

\correspondingauthor{Marco De Paoli}{\href{mailto:m.depaoli@utwente.nl}{m.depaoli@utwente.nl},\href{mailto:marco.de.paoli@tuwien.ac.at}{marco.de.paoli@tuwien.ac.at}}

\begin{keypoints} 
\item Convective mixing of carbon dioxide in geological formations follows different regimes
\item During a regime in which the flux is constant, mixing in 3D simulations is 13\% larger than in 2D simulations, well below previous estimates
\item A simple model can be employed to accurately reproduce the flow evolution for all the regimes and all the Rayleigh-Darcy numbers
\end{keypoints}
\justifying
\begin{abstract}
We perform large-scale numerical simulations of convection in 3D porous media at Rayleigh-Darcy numbers up to $\ra=8\times10^4$. To investigate the convective mixing of carbon dioxide (CO$_2$) in geological formations, we consider a semi-infinite domain, where the CO$_2$ concentration is constant at the top and no flux is prescribed at bottom. Convection begins with a diffusion-dominated phase, transitions to convection-driven solute finger growth, and ends with a shutdown stage as fingers reach the bottom boundary and the concentration in the system increases. For $\ra \ge 5\times10^3$, we observe a constant-flux regime with dissolution flux stabilizing at 0.019, approximately 13\% higher than in 2D estimates. Finally, we provide a simple and yet accurate physical model describing the mass of solute entering the system throughout the whole mixing process. These findings extend solutal convection insights to 3D and high-$\ra$, improving the reliability of tools predicting the long-term CO$_2$ dynamics in the subsurface.
\end{abstract}
\section*{Plain Language Summary} 
The geological sequestration of carbon dioxide (CO$_2$) consists of injecting large amounts of CO$_2$ into underground formations with the aim of permanent storage. This process is key in reducing greenhouse gas emissions in the atmosphere and in supporting energy transition. After injection, CO$_2$ combines with the resident fluid (brine) and remains stably trapped in the formation, preventing leakage to the atmosphere. The mixing process of CO$_2$ and brine can take place over hundreds or thousands of years and originates flows driven by density that, in turn, influence mixing. Predicting the evolution of this process is crucial in the design of injection strategies. Field and laboratory measurements are challenging due to the inaccessibility of underground sites and the time scale of the flow processes. Here we employ massive numerical simulations to predict the underground CO$_2$ dynamics, which depends on the fluid properties, on the rock properties and on the morphology of the formation. We systematically investigate the flow dynamics in three-dimensional systems, and provide a robust quantification of the differences occurring with respect to ideal two-dimensional systems. Finally, we derive a simple, reliable and accurate physical model to predict, design and control the post-injection dynamics of CO$_2$.

\section{Introduction}
Natural convection occurs when a flow is driven by density differences that exist within a fluid domain subject to gravity.
Convective flows occurring within a porous medium and characterized by density variations induced by heat or solute, are particularly relevant to geophysical processes.
The formation of sea ice, for instance, is controlled by a convective flow \cite{worster2019mushy}, driven by salt- and temperature-induced density variations, and it occurs within a newly formed and porous icy matrix, defined mushy layer \cite{feltham2006sea,worster2019mushy,wells2019mushy}. 
In dry salt lakes, subsurface convective processes are responsible for the vertical transport of salt, a fundamental mechanism in arid regions, leaving superficial polygonally-shaped crusts as signature \cite{lasser2021stability,lasser2023salt}. 
Recently, the problem of convection in porous media has become pivotal to address the challenge of carbon dioxide (CO$_2$) sequestration \cite{huppert2014fluid}, a key process required to contribute to the energy transition.

With the aim of permanent storage, supercritical CO$_2$ is injected in underground geological formations (e.g., saline aquifers), which may be idealized as porous regions filled with brine and confined by two low-permeability layers, acting as hydrodynamics barriers to the flow \cite{neufeld2010convective,depaoli2021influence}.
In these conditions, supercritical CO$_2$ is lighter than the resident fluid and after injection accumulates underneath the upper confining layer \cite{emami2015convective}. 
This situation is potentially critical, as a fracture of this natural barrier may allow CO$_2$ to escape to the upper geological layers and eventually to the atmosphere. 
However, CO$_2$ is partially miscible with brine, and the resulting mixture (CO$_2$+brine) is heavier than both starting fluids.
The process of accumulation of this heavy mixture at the boundary between the fluid layers is key, as it promotes the formation of flow instabilities (called plumes or fingers) that, due to convection, increase the mixing rate of CO$_2$ in brine \cite{ennis2005onset,xu2006convective} and reduce the risk of leakage.
Therefore, predicting the mixing rate as a function of fluid properties, rock properties and formation morphology is the key to accurately address the process of geological carbon sequestration \cite{ulloa2022energetics,letelier2023scaling}.

The dynamics of the CO$_2$-brine mixing process is characterized by three main phases \cite{slim2013dissolution,slim2014solutal}: i) initially, solute transport is limited to the fluid-fluid interface, and it is controlled by diffusion; ii) then, when the dense, CO$_2$-rich fluid layer becomes sufficiently thick, it becomes unstable and promotes the formation of small fingers; iii) these fingers grow vertically and eventually merge with neighboring plumes generating larger and persistent structures in a statistically-steady fashion; iv) finally, when  the domain is saturated with the CO$_2$-rich fluid carried downward by the megaplumes, the mixing of CO$_2$ slows progressively down \cite{hewitt2013convective}.
The single dimensionless governing parameter is the Rayleigh-Darcy number $\ra$, which expresses the ratio of the strength of driving (buoyancy) to dissipative (viscosity and molecular diffusion) mechanisms.
It contains the fluids, medium and formation properties, and it may span over few decades, namely from $O(10)$ \cite{EmamiMeybodi2017} to $O(10^5)$ \cite{hu2023effects}.
Despite the highly non-linear and transient character described, this system has been accurately characterized numerically in two dimensions \cite{emami2015convective,hew20,depaoli23}.
The three-dimensional problem, in contrast, is still partially unexplored, due to the huge computational cost required to accurately resolve the flow at large $\ra$. 
With the exception of few works at moderate $\ra$ \cite{Liyanage2024,croccolo2024,sin2024three,imuetinyan2024transparent}, the opacity of the media represents an intrinsic limitation for the laboratory experiments, preventing from obtaining simultaneous measurements of flow and solute. 
In this work, we aim precisely at this gap.

We perform high-resolution, large-scale simulations of solute convection in two- and three-dimensional porous media at unprecedented Rayleigh-Darcy numbers, namely up to $\ra=8\times10^4$. 
Using this unique database, which we make available \cite{databasethiswork}, we provide a detailed characterization of the flow dynamics.
We analyze the flow in terms of the dissolution rate of solute and flow structures near the fluid-fluid interface.
First, we confirm that in two-dimensional systems the dissolution rate during the constant-flux regime (phase iii described above) agrees with previous findings \cite{hesse2012phd,hewitt2013convective,slim2014solutal,depaoli2017solute,wen_2018}.
Then we extend our simulations to 3D flows, and we identify a remarkable difference with respect to previous works \cite{pau2010high}, suggesting that in 3D the dissolution rate is not 25\% larger than in 2D, but only 13.5\%.
Finally, we present a physical model that accurately captures the mixing dynamics, and can be used to design, predict and control the CO$_2$ post-injection dynamics.

\section{Methodology}\label{sec:meth}

\begin{figure}
\centering
\includegraphics[width=0.95\columnwidth]{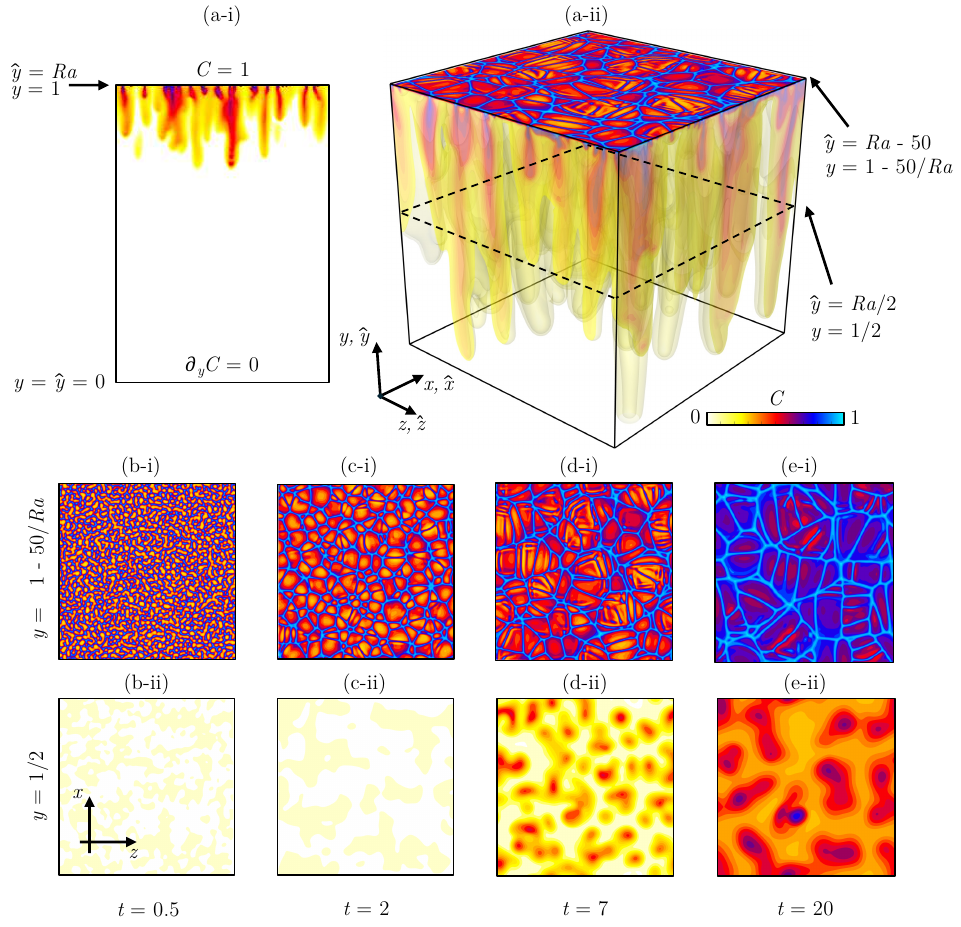}
\caption{\label{fig:config} 
Flow configuration (simulation B6, $\ra = 5 \times 10^3$, see Tab.~\textcolor{red}{S1}).
(a-i)~Concentration field at time $t = 2.0$, with indication of the boundary conditions and location of the domain wall (convective - $y$ - and diffusive - $\widehat{y}$ - units, see Sec. \ref{sec:dimeq34}). 
(a-ii)~Concentration distribution at $t = 7$. 
(b-e)~Concentration distribution over horizontal $x,z$ planes located at $y=1-50/\ra$ (or $\widehat{y}=\ra - 50$, panels i) and $y=1/2$ (or $\widehat{y}=\ra/2$, panels ii), at different times (indicated below panels~ii).
}
\end{figure}

\subsection{Problem formulation}\label{sec:prob}
We consider homogeneous, isotropic and fluid-saturated porous media characterized by permeability $\kappa$ and porosity $\phi$. 
We indicate with $x^*,z^*$ the horizontal directions and with $y^*$ the vertical direction, perpendicular to the horizontal boundaries and aligned with gravity~$\mathbf g$ (see Figs.~\ref{fig:config}a-i,a-ii).
Assuming the validity of the Boussinesq approximation \cite{landman2007heat,zonta2018stably}, the flow is incompressible and described by the continuity and the Darcy equations:
\begin{equation}
\nabla^{*}\cdot\mathbf{u}^{*}=0\quad,\quad
\mathbf{u}^{*}=-\frac{\kappa}{\mu}\left(\nabla^{*} P^{*}+\rho^{*}g \mathbf{j}\right) \text{  ,} 
\label{eq:eqadim2}
\end{equation}
with $\mu$ the fluid dynamic viscosity (constant), $\mathbf{u}^{*}=(u^{*},v^{*},w^{*})$ the volume-averaged velocity field, $P^{*}$ the pressure, and $\mathbf{j}$ the vertical unit vector ($^{*}$ indicates dimensional quantities).
The flow is driven by variations in fluid density, $\rho^{*}$, induced by the presence of a solute (CO$_2$) that is quantified by the concentration field $C^{*}$, with $0\le C^* \le C^{*}_\text{max}$.
We consider the density to be a linear function of the solute concentration:
\begin{equation}
\rho^{*}(C^{*})=\rho^{*}(0)+\Delta\rho^{*}\frac{C^{*}}{C^{*}_\text{max}}
\text{  ,} 
\label{eq:eqadim4}
\end{equation}
with $\Delta\rho^{*}=\rho^{*}(C^{*}_\text{max})-\rho^{*}(0)$.
The concentration field evolves according to the advection-diffusion equation \cite{nield17}
\begin{equation}
\phi\frac{\partial C^{*}}{\partial t^{*}}+\nabla^{*} \cdot(\mathbf{u}^{*}C^{*}-\phi D \nabla^{*} C^{*})=0 \text{  ,} 
\label{eq:eqadim3}
\end{equation}
where $t^{*}$ is time and $D$ is the solute diffusivity (considered constant).

The walls are impermeable to the fluid, at the upper boundary the concentration is  $C^{*}(y^*=H^*)=C^{*}_\text{max}$, while at the lower boundary a no-flux condition is applied.
The fluid is initially still and depleted of solute. 
Additional details are provided in Text~\textcolor{red}{S1}.

\subsection{Dimensionless equations}\label{sec:dimeq34}
Natural flow scales for convective flows are the buoyancy velocity, $\mathcal{U}^{*}=g \Delta \rho^{*} \kappa / \mu$, and the domain height, $H^{*}$.
Employing the set of dimensionless variables:
\begin{equation}
C=\frac{C^{*}}{C^{*}_\text{max}},\quad x=\frac{x^{*}}{H^{*}},\quad \mathbf{u}=\frac{\mathbf{u}^{*}}{\mathcal{U}^{*}},
\label{eq:eqadim5aaa}
\end{equation}
\begin{equation}
t=\frac{t^{*}}{\phi H^{*}/\mathcal{U}^{*}},\quad p=\frac{p^{*}}{\Delta \rho^{*}gH^{*}} ,
\label{eq:eqadim5}
\end{equation}
and using the reduced pressure $p^{*}=P^*+\rho^*(0)gy^*$, the dimensionless form of  Eqs.~\eqref{eq:eqadim2},\eqref{eq:eqadim3} is:
\begin{equation}
\nabla\cdot{\bf u}=0\quad,\quad{\bf u}=-\left( \nabla p + C {\bf j}\right) ,
\label{eq:equ1bis3}
\end{equation}
\begin{equation}
\label{eq:equ1bis1}
\frac{\partial C}{\partial t}+\nabla \cdot \left( {\bf u} C -\frac{1}{\ra} \nabla C\right)=0 ,
\end{equation}
where  
\begin{equation}
\ra=\frac{g \Delta \rho^{*} \kappa H^{*} }{ \phi D \mu}=\frac{\mathcal{U}^{*} H^{*}}{\phi D}
\label{eq:rada}
\end{equation}
is the Rayleigh-Darcy number (hereinafter indicated as Rayleigh number).
The problem is controlled by three dimensionless parameters: $\ra$ and the domain aspect ratios, $L_x=L^*_x/H^{*}$ and $L_z=L^*_z/H^{*}$. 
We considered domains with square  cross section ($L_x=L_z=L$).

To compare systems having different $\ra$, a convenient reference length scale should be independent of the domain height \cite{fu2013pattern}.
Taking as reference length scale the distance over which advection and diffusion balance, defined as $\ell^*=\phi D/\mathcal{U}^*$ \cite{fu2013pattern,slim2014solutal}, we obtain the dimensionless variables in diffusive units (indicated with $\widehat{\cdot}$ ):
\begin{equation}
\widehat{x}=\frac{x^{*}}{\ell^{*}} \quad,\quad  
\widehat{t}=\frac{t^{*}}{\phi \ell^{*}/\mathcal{U}^{*}}\quad ,\quad 
\widehat{p}=\frac{p^{*}}{\Delta \rho^{*}g\ell^{*}} ,
\label{eq:eqadimv2}
\end{equation}
and the Rayleigh number can be interpreted as a dimensionless domain height, $\ra = H^*/\ell^*$.
Note that the dimensionless form of other quantities in Eqs.~\eqref{eq:eqadim5aaa}-\eqref{eq:eqadim5} remains unchanged.
Both dimensionless forms of the flow variables will be employed to discuss the results. 

Eqs.~\eqref{eq:equ1bis3}-\eqref{eq:equ1bis1} are solved numerically with the aid of a second-order finite-difference in-house solver \cite{pirozzoli21,depaoli22} (see also \citeA{kim_85,orlandi_00,pirozzoli_14,pirozzoli_17} and Texts~\textcolor{red}{S1-S3} of the Supporting Information).

\section{Results}\label{sec:theo}
In this work, we provide for the first time a systematic comparison of the 2D and 3D dissolution dynamics at unprecedented high $\ra$, namely $10^2\le\ra\le8\times10^4$.
The simulations details are given in Tab.~\textcolor{red}{S2}.
We present the results in terms of mixing, quantified by the flux of the solute through the top wall (Sec.~\ref{sec:flux}) and the cumulative amount of dissolved solute (Sec.~\ref{sec:mass}), and we relate the mixing dynamics to the flow structures.

\begin{figure}
\centering
\includegraphics[width=0.95\columnwidth]{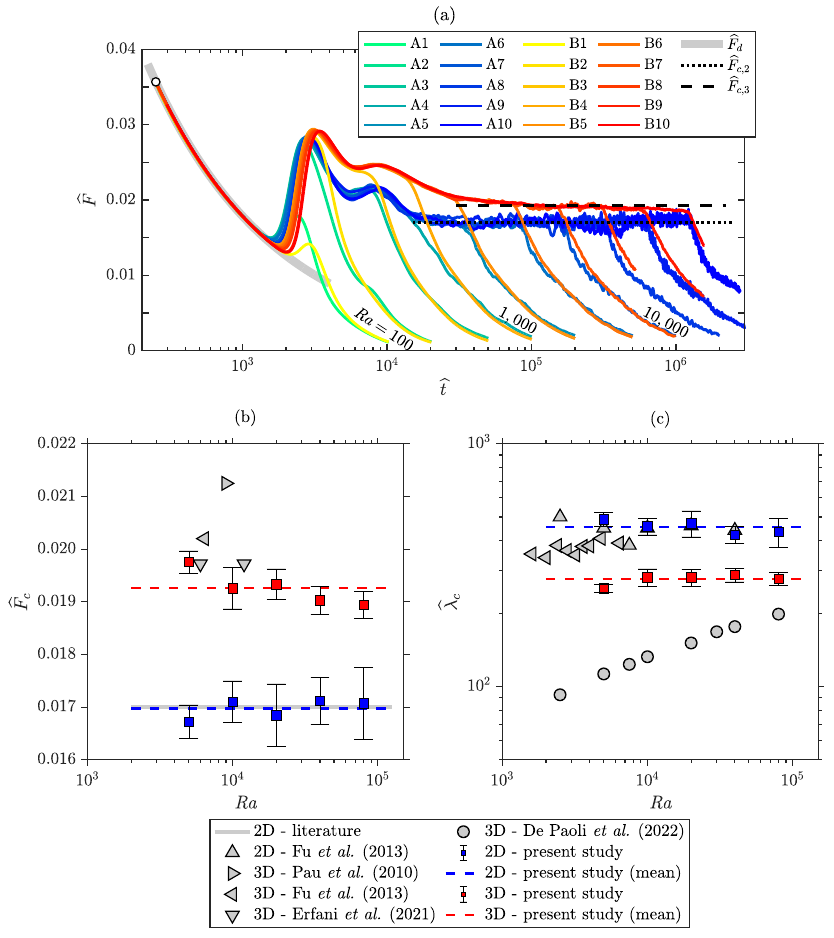}
\caption{
(a)~Flux $\widehat{F}$, defined in~\eqref{eq:eqflux1}, obtained in 2D (A\#) and 3D (B\#) simulations (details in Tab.~\textcolor{red}{S2}).
Results are shown from time $\widehat{t}=250$ ($\circ$) to $\widehat{t}\approx100\ra$.
The diffusive flux $\widehat{F}_d$, Eq.~\eqref{eq:eqflux2}, and the mean value in the constant-flux regime (2D, $\widehat{F}_{c,2}$ in Eq.~\eqref{eq:eqflux3} and 3D, $\widehat{F}_{c,3}$ in Eq.~\eqref{eq:eqflux4}) are also shown.
(b)~Time-averaged flux during the constant-flux regime for $\ra\ge5\times10^3$ ($\widehat{F}_c$, squares) and corresponding standard deviation ($\sigma(\widehat{F})$, error bar). The ensemble averaged values (dashed lines) are also shown.
Results are compared against literature for 2D (see Sec.~\ref{sec:flux2}) and 3D \cite{pau2010high,fu2013pattern,erfani2021dynamics}.
(c)~Length scales of the flow structures ($\widehat{\lambda}_c$ and corresponding deviation) in the vicinity of the top boundary and averaged during the constant-flux regime.
Results are compared against measurements by \citeA{fu2013pattern} (one-sided) and \cite{depaoli22} (two-sided, Rayleigh-B\'enard-Darcy).
}\label{fig:flux1} 
\end{figure}

\subsection{Flux of solute dissolved}\label{sec:flux}
The amount of solute that enters the domain per unit of surface area and time is quantified by the dimensionless flux $\widehat{F}(t)$, defined as:
\begin{equation}
\widehat{F}(t) = \int_{\widehat{A}} \frac{\partial C}{\partial \widehat{y}} \bigg\rvert_{\widehat{y}=1}  \text{ d}\widehat{A},
\label{eq:eqflux1}
\end{equation}
where $\widehat{A}=L_x^*/\ell^*$ in 2D and $\widehat{A}=L_x^*L_z^*/(\ell^*)^2$ in 3D. 
The evolution of this response parameter, representative of the actual dissolution dynamics of the system, is reported as a function of the diffusive time $\widehat{t}$ in Fig.~\ref{fig:flux1}(a).

The flow dynamics in 2D and 3D systems is qualitatively similar (see also Fig.~\textcolor{red}{S4}). 
As described in previous 2D studies \cite{slim2013dissolution,hewitt2013convective,slim2014solutal,depaoli2017solute,wen_2018}, the evolution is characterized by several regimes: (i)~the system is first controlled by diffusion, as the initial condition prescribes $\mathbf{u}=0$; the fluid density at the upper boundary layer increases as a result of the concentration contrast between the bulk of the domain and the boundary value, building up a layer of heavy fluid that eventually becomes unstable \cite{slim2014solutal}; (ii)~later fingers form (Fig.~\textcolor{red}{S4}a), the flux grows and fingers merge; (iii)~the system attains a statistically-steady state, with large and persistent plumes dominating the bulk of the flow and extending vertically (Fig.~\textcolor{red}{S4}b); (iv)~finally, after the fingers reach the lower boundary, the system progressively saturates (Fig.~\textcolor{red}{S4}c): the concentration contrast between the bulk of the domain and the top boundary value diminishes, and it is increasingly harder for the solute to enter the system. 
We will analyze the dynamics in regimes (i), (iii) and (iv) in Secs.~\ref{sec:flux1}, \ref{sec:flux2} and \ref{sec:flux3}, respectively.

\subsubsection{Diffusive phase}\label{sec:flux1}
Initially, the flow dynamics is marked by the absence of convection ($\mathbf{u}=0$).
Eq.~\eqref{eq:equ1bis1} has a self-similar solution, Eq.~(\textcolor{red}{S1}), that combined with Eq.~\eqref{eq:eqflux1} allows to determine the evolution of the flux \cite{slim2014solutal,depaoli2017solute}: 
\begin{equation}
\widehat{F}_d(\widehat{t}) = \frac{1}{\sqrt{\pi\widehat{t}}}.
\label{eq:eqflux2}
\end{equation}
We observe from Fig.~\ref{fig:flux1}(a) that the evolution of the system, both in 2D and in 3D, follows Eq.~\eqref{eq:eqflux2} up to $\widehat{t}\approx10^3$.
After the fingers' formation, the flux grows departing from the diffusive state.

\subsubsection{Constant flux phase}\label{sec:flux2}
An initial flux growth is followed by the continuous merging of newly formed fingers into large descending plumes that are persistent in time and space (see Movie~\textcolor{red}{S1} and relative discussion in the Supporting Information). 
The dynamics of the system is statistically steady, and this phase is labeled as \textit{constant-flux regime}.
The transition from the diffusive to the constant-flux regime occurs approximately for $2\times10^3 \le \widehat{t} \le \widehat{t}_1$, where $\widehat{t}_1=2\times 10^4$ in 2D and $4\times 10^4$ in 3D, respectively. 
Predicting the time at which this transition occurs is challenging, as it depends on the amplitude of the initial perturbation applied and the $\ra$ considered (see \citeA{slim2010onset,elenius2012time,riaz2006onset,riaz2014carbon} and Text~\textcolor{red}{S1}). 
In addition, small-$\ra$ simulations (A1-A4 and B1-B5) do not reach the constant-flux stage, as the domain rapidly saturates.  The dimensionality of the system also plays a key role: the process of merging of the fingers is different in 3D compared to the 2D case, provided the additional degree of freedom available, and for any $\ra$ the flux is larger in 3D than in 2D, as also observed in Rayleigh-B\'enard-Darcy systems \cite{hew14,pirozzoli21,depaoli24}.

An accurate estimate of $\widehat{F}$ during the constant-flux regime is essential, as it is used as a subgrid-scale parameter for large-scale models based on multiphase gravity currents, commonly used to predict the long-term post-injection dynamics \cite{gasda2011vertically,macminn2012spreading,macminn2013buoyant,hidalgo2013dynamics,depaoli2021influence}.
In the last two decades, several numerical measurements have been provided in 2D, but very few results are available in 3D, also due to the prohibitive computational costs involved. 
Here we provide for the first time a systematic direct comparison of 2D and 3D flows, and we quantify the dissolution during the constant-flux regime at high $\ra$.

To determine the mean flux during the constant-flux regime, $\widehat{F}_c$, we only consider the simulations that achieve the constant flux condition, namely simulations A6$-$A10 in 2D and simulations B6$-$B10 in 3D (see Fig.~\ref{fig:flux1}a).
The constant-flux regime takes place later in 3D compared to the 2D case.
Therefore, we chose different time intervals to determine the time-averaged flux, namely $\widehat{t}_{c,1}\le \widehat{t} \le \widehat{t}_{c,2}$ with $\widehat{t}_{c,1} = 2\times 10^4$ and $\widehat{t}_{c,2} =16\ra$ in 2D, and $\widehat{t}_{c,1} = 4\times10^4$ and $\widehat{t}_{c,2} =14\ra$ in 3D.
The upper and lower limits are determined by looking at the evolution of $\widehat{F}$.
Note that while the lower limit is constant when defined in diffusive units ($\widehat{t}$), the upper limit is constant in convective units ($t$).
This is justified by the initial self-similar character of the flow, the behavior of which is independent of the value of $\ra$ until the fingers impact on the lower wall ($t\approx7-8$) \cite{hewitt2013convective,slim2014solutal,depaoli2017solute,wen_2018}.

We show in Fig.~\ref{fig:flux1}(b) the time-averaged values of flux during the constant-flux regime, $\widehat{F}_c$, provided with the standard deviation $\sigma(\widehat{F})$ within the same interval (values listed in Tab.~\textcolor{red}{S2}).
In 2D, present results are in excellent agreement with previous works \cite{pau2010high,hesse2012phd,hewitt2013convective,slim2014solutal,depaoli2017solute,wen_2018,erfani2021dynamics,depaoli2024afid} who reported $\approx 0.017$ (gray solid line in Fig.~\ref{fig:flux1}b).
The mean value ($\pm$ deviation) obtained here is
\begin{equation}
\widehat{F}_{c,2}=0.01697\pm0.00018,
\label{eq:eqflux3}
\end{equation}
indicated in Fig.~\ref{fig:flux1}(b) by a blue dashed line.
Both 0.01697 and 0.0017 fall within the error bars displayed.

In the 3D case, the flux exhibits a weakly decreasing trend, with the constant flux value diminishing from 0.01975 (B6) to 0.01894 (B10), and a mean value
\begin{equation}
\widehat{F}_{c,3}=0.01926\pm 0.00032,
\label{eq:eqflux4}
\end{equation}
shown in Fig.~\ref{fig:flux1}(b) as a red dashed line.
In 3D (see, e.g., simulation B6), the accuracy of Eq.~\eqref{eq:eqflux4} in describing the mean flux is lower than in the 2D case.
However, we observe that our results are in agreement with previous findings \cite{fu2013pattern,erfani2021dynamics}, with the exception of the pioneering work \citeA{pau2010high}, who reported a flux being 25\% larger than in 2D, while in our case we obtain a difference of
\begin{equation}
100\left(\frac{\widehat{F}_{c,3}}{\widehat{F}_{c,2}}-1\right)=13,50\% .
\label{eq:eqflux5}
\end{equation}
The implications of these findings are considerable, as \citeA{pau2010high} overestimate by a factor $\approx$2 the increased efficiency of the 3D transport mechanisms compared to the 2D ones. 
In Eq.~\eqref{eq:eqflux5}, when replacing $\widehat{F}_{c,3}$ with the minimum and maximum values of $\widehat{F}$ during the constant-flux regime for simulation B10, we obtain 8.99\% and 17.83\%, respectively, still well below the estimate of \citeA{pau2010high}.

To quantify size and numbers of fingers close to the upper boundary, we compute the power-averaged mean wavenumber $k$ \cite{de2004miscible}:
\begin{equation}
    k(t) = \frac{\int_0^{k_N} k_h |\tilde{C}(k^*,t)|^2\text{ d}k^*}{\int_0^{k_N}  |\tilde{C}(k^*,t)|^2\text{ d}k^*},
    \label{eq:pawn}
\end{equation}
where $k_h$ is the horizontal wavenumber \cite[$0\le k_h\le k_N$ with $k_h=k_x$ in 2D, and $k_h=\sqrt{k_x^2+k_z^2}$ in 3D, and $k_N$ the maximum wavenumber]{fu2013pattern,hew14}, while $|\tilde{C}|$ is computed by taking the Fourier transform of the concentration field in horizontal direction and is used to obtain the power spectrum density $|\tilde{C}(k^*,t)|^2$.
The corresponding wavelength $\lambda = 1 / k$ is representative of the average cell size, or plume diameter \cite{fu2013pattern}.
The behavior of $k$ in time (not shown) follows the time-dependent evolution of the dissolution flux.
In both 2D and 3D, the wavenumber evolves independently of $\ra$ until the system enters the shutdown regime.
During the constant-flux regime, the wavenumber is also constant.
To compare the dynamics at different $\ra$, the representative size of the structures expressed in diffusive units, $\widehat{\lambda}=\ra/k$, is averaged during the constant-flux regime, and we indicate the corresponding value with $\widehat{\lambda}_c$.
The values relative to each simulation (reported in Tab.~\textcolor{red}{S2}) indicate that the cell size is larger in 2D than in 3D, and correspond to $\widehat{\lambda}_{c,2}\approx455$ and $\widehat{\lambda}_{c,3}\approx277$, respectively.
During the constant-flux regime there is no external length scale controlling the system, and the characteristic scale of the flow is the length over which advection and diffusion balance, $\ell^*$.
We observe in Fig.~\ref{fig:flux1}(c) that $\widehat{\lambda}_c\sim \ra^{0}$ as predicted by \citeA{fu2013pattern}.
We also observe that the present 2D results are in excellent agreement with the literature.
In 3D, results are still independent of $\ra$, but have a different value compared to \citeA{fu2013pattern}, possibly because their definition of the average cell size is not based on the mean wavenumber, but it is obtained from the number of plumes present in the system.
The scaling $\widehat{\lambda}_c\sim \ra^{0}$ is observed both in 2D and 3D flows, and we can finally confirm the applicability at large $\ra$ of the scaling previously proposed by \citeA{fu2013pattern}, who highlighted the need of 3D measurements at high-$\ra$.

As a reference, we also report the mean wavelength obtained in Rayleigh-B\'enard-Darcy configuration \cite{depaoli22} (circles in Fig.~\ref{fig:flux1}c), where $\widehat{\lambda}\sim\ra^{0.19}$ and the flow morphology differs from the present case.
This is possibly due to the more complex relationship between the response parameter (Nusselt number) and $\ra$ than in one-sided convection \cite{depaoli24}.

\subsubsection{Convective shutdown phase}\label{sec:flux3}
When a sufficient amount of solute has entered the system, the domain homogenizes and the bulk concentration increases progressively (see Fig.~\textcolor{red}{S4}c), reducing the fluid density contrast between the upper boundary and the bulk.
As a result, the effective driving force of the flow is reduced, as well as the amount of solute entering the system, per unit time. 
This mechanism is defined as \textit{shut-down of convection} \cite{hewitt2013convective}. 
In 2D, it has been accurately described by so-called ``box models"  \cite{hewitt2013convective,slim2014solutal}.
Here we will employ the approach of \citeA{slim2014solutal}, which relies on the approximation of well-mixed bulk (also verified in the 3D case, see Fig.~\textcolor{red}{S4}c).
Therefore, this model is designed to describe well the flow behavior during the late phases, while the accuracy during the transition from the constant-flux to the shutdown regime is lower.

Assuming a uniform concentration field far from the upper wall, one can derive an exact form for the evolution of the bulk concentration, which is related to: (a)~the critical Rayleigh number \cite[$\ra_{c}=4\pi^2$]{horton1945convection,lapwood1948convection}, (b)~the beginning of the shutdown regime $\left(\widehat{t}_{sd}\right)$ and (c)~a fitting parameter related to the bulk concentration $(c_0)$.
Combining this expression of the bulk concentration with definition~\eqref{eq:eqflux1} and integrating Eq.~\eqref{eq:equ1bis1} over the volume, we obtain a prediction for the evolution of the shutdown flux:
\begin{equation}
    \widehat{F}_{sd}(t) = \frac{\ra_{cr}(1-c_0)^2}{\left[\left(1-c_0\right)\left(\widehat{t}/\ra-t_{sd}\right)+\ra_{cr}\right]^2}.
    \label{eq:fluxsh_0}
\end{equation}
We refer to \citeA{slim2014solutal} for further details on the model's derivation.

In 2D, we find that the best-fitting parameters are $\ra_{cr}=31.5$, $t_{sd}=16$ and $c_0=0.2716$, analogous to the results reported by \citeA{slim2014solutal}.
Comparison of the model predictions against the numerical results (not shown here) indicates that Eq.~\eqref{eq:fluxsh_0} approximates the flux also very well for the present simulations, especially for large $\ra$ and far from the transition to shutdown. 
Similarly, we found that the model parameters that are suitable to describe the 3D system are $\ra_{cr}=27.0$, $t_{sd}=14$ and $c_0=0.2697$ (further details on the constraints of these parameters are provided in Sec.~\ref{sec:mass2}).
One can observe in Fig.~\textcolor{red}{S5} that also in the 3D case the predictions of the model~\eqref{eq:fluxsh_0} are accurate.

\subsection{Volume of solute dissolved}\label{sec:mass}

Although the flux $\widehat{F}\left(\widehat{t}\right)$ is representative of the instantaneous flow configuration, determining the cumulative amount of flux dissolved is also crucial, as it allows us to estimate, for instance, how long it takes for the system to dissolve a given volume of solute.
In the frame of CO$_2$ storage, this is key to determine whether or not a formation is safe: if the time taken to dissolve the injected fluid is too long, then it is possible that a fracture in the upper low-permeability layer confining the buoyant CO$_2$ may lead to a leakage to the upper layers before a large proportion of CO$_2$ has mixed.
Here we quantify the amount of solute mixed via the parameter
\begin{equation}
    \widehat{M}\left(\widehat{t}\right)=\int_0^{\widehat{t}}\widehat{F}(\tau)\text{ d}\tau,
    \label{eq:vol1}
\end{equation}
and the corresponding quantity in convective units reads $M=\widehat{M}/\ra$.
Note that $M$ is also representative of the degree of mixing \cite[$M=0$ means solute-free, while the volume is fully saturated for $M=1$]{jha2011quantifying}.
Correspondingly, we find that $0\le\widehat{M}\le\ra$.

The volume dissolved expressed in convective units is reported in Fig.~\textcolor{red}{S6}(a) as a function of time $\widehat{t}$ (diffusive units). 
This choice of dimensionless variables allows us to distinguish the behavior at different $\ra$ and also to directly compare the evolution of 2D and 3D flows. 
Provided that the flux in 3D is larger than in 2D (see Fig.~\ref{fig:flux1}a), $M$ grows also faster in 3D than in 2D. 
A small difference may occur at very early times ($\widehat{t}<2\times10^3$), due to an earlier onset of convection in 2D compared to 3D: this is responsible for the slightly larger values of $M$ observed in Fig.~\textcolor{red}{S6}(a) for A1-A3 compared to B1-B3.

In Fig.~\textcolor{red}{S6}(b) we evaluate the accuracy of the model in describing the 3D results. 
Integrating Eq.~\eqref{eq:fluxsh_0} as in \eqref{eq:vol1}, we obtain: 
\begin{equation}
\widehat{M}_{sd}\left(t\right)=\widehat{M}\left(t_0\right)+\int_{t_0}^{t}\widehat{F}_{sd}(\tau)\text{ d}\tau,
    \label{eq:vol2}
\end{equation}
where we considered as a starting point $t_0=14$ and $\widehat{M}\left(t_0\right)$, with $\widehat{M}\left(t_0\right)$ is computed numerically as in \eqref{eq:vol1}.
The point $(t_0,\widehat{M}\left(t_0\right))$ is marked by a symbol ($\circ$) in Fig.~\textcolor{red}{S6}(b).
Choosing the value measured from the simulations as a starting point allows us to evaluate the accuracy of the model in the shutdown phase, which appears to be excellent for all $\ra$ considered.
In the following, we will provide a formulation that is independent of the starting point.

\subsubsection{Model for the volume dissolved across all phases}\label{sec:mass2}
We focus on the high-$\ra$ behavior in which all the regimes are present, and therefore we only consider for the model validation the simulations B5$-$B10, which we report in Fig.~\ref{fig:6}(a-b). 
We split the flow evolution into three main regimes \cite{slim2014solutal}, discussed in Sec.~\ref{sec:flux}: diffusive, constant-flux and shutdown.
The unified piecewise model reads: 
\begin{equation}
    \widehat{M}\left(\widehat{t}\right)=
    \begin{cases}
        &\frac{2\sqrt{\widehat{t}}}{\sqrt{\pi}} \quad,\quad \text{for } \widehat{t}<\widehat{t}_1\\
        &\frac{2\sqrt{\widehat{t_1}}}{\sqrt{\pi}} + \widehat{F}_{c}\left(\widehat{t}-\widehat{t}_1\right) \quad, \quad \text{for } \widehat{t}_1 \le \widehat{t}< \widehat{t}_2 \\
        &\frac{2\sqrt{\widehat{t_1}}}{\sqrt{\pi}} + \widehat{F}_{c}\left(\widehat{t}_2-\widehat{t}_1\right) + 
        \frac{\ra(1-c_0)^2\left(\widehat{t}/\ra-t_{sd}\right)}{(1-c_0)\left(\widehat{t}/\ra-t_{sd}\right)+\ra_{cr}} 
        \quad, \quad \text{for }  \widehat{t}\ge \widehat{t}_2
    \end{cases}
    \label{eq:fluxmodel1}
\end{equation}
with $\widehat{t}_1=2\times 10^3$, $\widehat{t}_2=t_{sd}\ra$ and $\widehat{F}_{c}$ defined as in Sec.~\ref{sec:flux2}.
This simple model is obtained from integration in time of Eqs.~\eqref{eq:eqflux2}, \eqref{eq:eqflux4} and \eqref{eq:fluxsh_0} (model parameters employed are summarized in Tab.~\textcolor{red}{S3}). 
Note that $\widehat{M}(\widehat{t}\to\infty)=\ra$, assuming that a full saturation is achieved asymptotically.
This condition is achieved for 
\begin{equation}
c_0=t_{sd}\widehat{F}_{c} \quad\text{ and }\quad \widehat{t}_1=\left[2/\left(\sqrt{\pi}\widehat{F}_{c}\right)\right]^2,
\end{equation}
with $t_{sd}$ and $\widehat{F}_c$ measured from the numerical simulations. 
As a result of these additional constraints, the only fitting parameter is $\ra_{cr}$.

We observe in Fig.~\ref{fig:6}(a-b) that the model captures very well the dissolution dynamics during the whole flow evolution. 
The transition between two regimes is marked by symbols (diamond for $\widehat{t}_1$ - transition form diffusive to constant-flux -  and bullets for $\widehat{t}_2$ - constant-flux to shutdown).
The predictions of the model are excellent in the short term and also on the long term, see Figs.~\ref{fig:6}(a) and~\ref{fig:6}(b), respectively.
The availability of such a simple and accurate model is key to deliver long-term predictions on the flow evolution, which will be discussed in Sec.~\ref{sec:appl}.

\section{Application to CO$_2$ sequestration}\label{sec:appl}
We employ the model~\eqref{eq:fluxmodel1} to investigate the dissolution dynamics in terms of cumulative amount of solute dissolved (either volume or mass, since the flow is considered incompressible) over a wide range of times and $\ra$.

\begin{figure}
\vspace{-0.5cm}
\includegraphics[width=0.98\columnwidth]{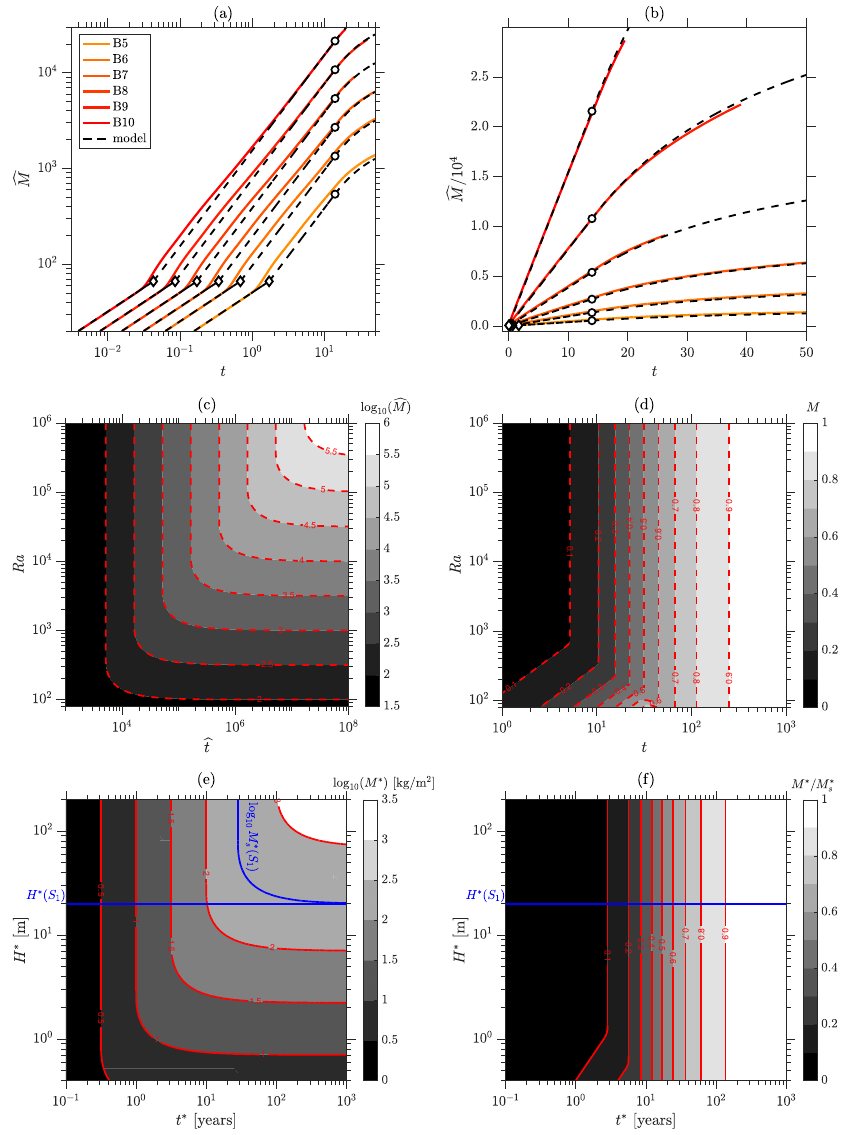}
\caption{
(a-b)~Volume of solute dissolved ($\widehat{M}$, diffusive units) in time ($t$, convective units).
Model \eqref{eq:fluxmodel1} is compared against 3D simulations with $\ra\ge2\times10^3$ (B5$-$B10). 
Symbols mark the transition between regimes (diamond $\widehat{t}_1$, bullet $\widehat{t}_2$).
(c-f) Model predictions for the mass of CO$_2$ dissolved per unit of surface area.
Dimensionless mass is shown as a function of $\ra$ and time, in diffusive (c) and convective units (d).
Mass as a function of domain height and time is shown in physical units (e) and normalized with respect to the maximum value, $M^*_s$ (f).
Scenario $S_1$ (Sleipner site, North Sea, see Sec.~\ref{sec:appl}) is indicated in blue. 
}\label{fig:6} 
\end{figure}

The volume of solute in the system, $\widehat{M}$, is presented in Fig.~\ref{fig:6}(c). 
For better visualization, $\log\widehat{M}$ is shown.
At very early times ($t\le\widehat{t}_1$), $\widehat{M}$ is independent of $\ra$. 
Afterwards, $\ra$ determines the dynamics: asymptotically the volume is $\widehat{M}\to\ra$ and the system saturates. 
A different interpretation is inferred when looking at Fig.~\ref{fig:6}(d), where the model results are shown in convective units ($t,M$).
For $\ra\le300$, the evolution is time-dependent and the constant-flux regime is never achieved. 
However, for larger $\ra$, the system becomes independent from $\ra$, as the shutdown phase is dominant in terms of amount of solute dissolved.
The value of $M$ displayed in convective units corresponds also to the fraction of volume saturated with solute. 
This suggests, for instance, that to saturate a volume corresponding to 50\% of the entire domain, $t\approx30$ is required for nearly all $\ra$.
In contrast, to achieve a 90\% saturation, a time $ t \approx 250$ is needed. 

This model can be employed, for instance, to identify the optimal time and $\ra$ required to dissolve a prescribed amount of solute $\widehat{M}$. 
We consider the Sleipner site in the North Sea (indicated by $S_1$).
The layer, considered homogeneous and isotropic, has permeability $\kappa=2.5\times 10^{-12}$~m$^2$ and porosity $\phi=0.375$. 
These data as well as the fluid properties later discussed are taken from \citeA{neufeld2010convective}.
The density contrast between the CO$_2$-saturated solution and brine is $\Delta\rho=10.5$~kg/m$^3$, and the fluid dynamic viscosity is $\mu=5.9\times10^{-4}$~Pa~s. 
The diffusion coefficient and the CO$_2$ solubility are $D=2\times 10^{-9}$~m$^2$/s and $C^*_\text{max}= 37.8$~kg/m$^3$ \cite{liu2024determination}, respectively.
This set of parameters fully defines the medium and the fluid properties, and determines the flow scales -- $\ell^*=1.718\times10^{-3}$~m and  $\mathcal{U}^*=4.364\times10^{-7}$~m/s -- and the corresponding average flow cell size near the CO$_2$-brine interface, $\lambda^*_c\approx0.5$~m.
In Fig.~\ref{fig:6}(e) we report the dissolved CO$_2$ mass per unit of surface area as a function of time ($t^*$) and layer depth ($H^*$).
In dimensional form, it is obtained as $M^*=\widehat{M}\phi D C^*_\text{max}\tau^*/\ell^*$, with $\tau^*=\phi\ell^*/\mathcal{U}^*$.
The maximum amount of solute that can be dissolved is achieved asymptotically and it corresponds to $M^*_s = C^*_\text{max}\phi H^*$.
For the Sleipner Site $H^*(S_1)=20$~m \cite{neufeld2010convective} (blue line in Fig.~\ref{fig:6}e), and we obtain $M^*_s(S_1)=283.5$~kg/m$^2$ and $\ra(S_1)=1.16\times10^4$.
Using this model we can also estimate the time required to dissolve a prescribed amount of CO$_2$: it takes approximately 10 years to mix 100~kg/m$^2$, and the same time is required for any formation with height $H^*\ge H^*(S_1)$.

The model presented can be also employed to determine the mixing state within the geological formation. 
For instance, for the fluid and the medium properties corresponding to $S_1$, it is possible to determine the fraction of solute dissolved relative to the maximum, $M^*/M^*_s$, see Fig.~\ref{fig:6}(f).
Considering the Sleipner Site ($H^*(S_1)$, blue line), it takes $<20$~years to achieve a 50\% dissolution, but $>100$~years to reach 90\%, due to the reduction of the strength of convective transport mechanisms during the shutdown phase.

\section{Conclusions and outlook}\label{sec:conc}
We use large-scale simulations to study solutal convection in porous media at unprecedented Rayleigh-Darcy numbers, $\ra$.
We provide, for the first time, a direct comparison of the dynamics of 2D and 3D systems over a wide range of $\ra$, $10^2\le\ra\le8\times10^4$, and we make this unique dataset available \cite{databasethiswork}.
The dissolution dynamics is characterized by the presence of several regimes, from an initial diffusion-dominated phase, to a quasi-steady state (constant-flux regime), and finally to the shutdown of convection due to the saturation of the domain.
A detailed analysis of the constant-flux regime reveals that only for \(\ra \ge 5\times10^3\) does a constant-flux regime occur, and that in 3D the dissolution rate is 13.5\% higher than in 2D.
We believe that this increase is the result of the non-trivial combination between (i)~the increased number of plumes \cite{pau2010high,fu2013pattern} and (ii)~the decreased advective flux per plume.
In fact, the increase in the number of plumes $(100(\lambda_{c,2}/\lambda_{c,3}-1)\approx60\%)$ is well beyond the increase in the flux ($13.5\%$).
This suggests that the vertical downward velocity in a 3D environment diminishes compared to a 2D environment, as also observed in other systems \cite{Boffetta2020,borgnino2021dimensional}.
In addition, the plumes geometry and coherence changes from 2D (``rectangular'' plumes) to 3D cases (``sheet-like'' plumes). 
This further complicates the picture and requires deeper analyses to provide a sound theoretical explanation of the results. 
These findings highlight the key role of the present 3D simulations, reporting considerably lower values of solute fluxes compared to previous estimates \cite{pau2010high}.
The implications of the present results are considerable since the constant flux is commonly used as a subgrid-scale  parameter in reservoir-scale simulations \cite{macminn2012spreading,neufeld2010convective}. 
We also proposed a simple physical model to describe the mass of solute entering the system throughout the whole mixing process.
Despite the model's simplicity, the predictions obtained are very accurate when compared against numerical 3D measurements, revealing the reliability of this modeling approach in describing the flow evolution across the different phases.
Finally, we exploit this model to capture the macroscopic dissolution dynamics during the process of CO$_2$ storage in the Sleipner Site (North Sea).

The results obtained are relative to homogeneous and isotropic porous media.
However, formations identified as possible sequestrations sites may be anisotropic \cite{depaoli2016influence,green2018steady}, heterogeneous \cite{chen2015diffuse,hidalgo2022advective} and/or fractured \cite{ulloa2025convection}, with possible non-trivial implications on the 3D mixing.
In addition, solute redistribution produced by mechanical dispersion \cite{liang2018,wen2018rayleigh,Tsinober2022,tsinober2023numerical,depaoli2024towards} may also impact the flow dynamics.
Depending on the chemical composition of the geological formation, flow-induced morphology modifications can also influence the dissolution dynamics \cite{Fu2015,hidalgo2015dissolution,miri2016salt}.
Investigating the role of these additional mechanisms is key to develop safe, reliable and effective CO$_2$ sequestration strategies \cite{wang2022analysis}.

\section*{Open Research Section}
The data of flux presented in this work are available at \citeA{databasethiswork}.
The data relative to the flux shown in Fig.~\ref{fig:flux1}(a) are available via \citeA{pau2010high,fu2013pattern,erfani2021dynamics,depaoli22}.

\acknowledgments
This project has received funding from the European Union's Horizon Europe research and innovation programme under the Marie Sklodowska-Curie grant agreement MEDIA  No.~101062123.
We acknowledge the EuroHPC Joint Undertaking for awarding the project GEOCOSE number EHPC-REG-2022R03-207 and for granting access to the EuroHPC supercomputer LUMI-C, hosted by the LUMI consortium (Finland).
The authors acknowledge TU Wien Bibliothek for financial support through its Open Access Funding Programme.
The anonymous referees, whose comments and suggestions helped to improve this manuscript, are also acknowledged.

\bibliography{bibliography}

\end{document}